\documentclass[a4paper,11pt]{revtex4}
\usepackage{graphicx,amssymb,amsmath,bm,latexsym,color,epsf}

\makeatletter
\@addtoreset{equation}{section}

\makeatother
\newcommand{\be}{\begin{equation}}
\newcommand{\ee}{\end{equation}}
\newcommand{\bea}{\begin{eqnarray}}
\newcommand{\eea}{\end{eqnarray}}

\def\Mpl{{M_{\rm Pl}}}
\begin{document}

\begin{titlepage}

\title{The Starobinsky model of inflation and Renormalizability
\\
\vskip 1cm
{\it \small Invited Contribution to the Starobinsky Memorial Volume, Springer Nature 2025}
}

\author{Roberto Percacci}
\email{percacci@sissa.it}
\affiliation{
SISSA, via Bonomea 265, I-34136 Trieste
and  INFN, Sezione di Trieste, Italy
}
\author{Gian Paolo Vacca}
\email{vacca@bo.infn.it}
\affiliation{INFN - Sezione di Bologna and DIFA,
via Irnerio 46, I-40126 Bologna}

\begin{abstract}
\vskip 1cm
{\centerline {\bf Abstract} }
\vskip 0.5cm

The Starobinsky model was born in a cosmological scenario where conformally coupled matter quantum field fluctuations on the vacuum drive a non trivial semiclassical energy momentum tensor quadratic in curvature. The presence of an unstable de Sitter solution of the semiclassical Einstein equations contributed to spread the idea that the early universe could have experienced an inflationary epoch.
Effective "$R + R^2$"  models of gravity have later gained much attention since their predictions are in very good agreement with the measurements of CMB data and tensor to scalar ratio bounds. In this paper we observe how the Starobinsky model can be  well approximated by the asymptotically free quadratic gravity on a part of a renormalization group trajectory (below some high UV scale) which is free from tachyonic instabilities,  if a definition of ``physical'' running is employed.
\end{abstract} 

\maketitle

\end{titlepage}
\newpage
\setcounter{page}{2}

\section{Introduction}
\label{sec:1}
One cannot underestimate the enormous contributions given by Alexei A. Starobinsky to the development of many aspects of early universe Cosmology. We shall discuss here some quantum features of gravity, which are specifically related to just one aspect of Alexei's work, known nowadays simply as {\it Starobinsky inflation}.

This scenario originated from his work~\cite{Starobinsky:1980te}, which  was very quantum and geometrical at the same time. Maybe picking up the path, started by A. Sakharov's idea of ``induced gravity''~\cite{Sakharov:1967pk}, Alexei studied a model of Einstein gravity in presence of conformally coupled massless fields, whose quantum vacuum fluctuations were the source of a non trivial effective energy momentum tensor, quadratic in the curvature, associated to the trace anomaly.  Solving the semiclassical Einstein equation of motion for the metric he was finding an (unstable) de Sitter solution, describing an early universe expanding phase.

Later it was understood that this kind of quadratic gravity models were interesting in cosmology~\cite{Starobinsky:1983zz,Vilenkin:1985md} for the inflationary paradigm and the consequent properties of particle production.

It is natural to think about gravitation interacting with quantum fields to be itself part of the quantum world, with propagating quantum fluctuations of its own. 
In an interacting theory like gravity, these quantum fluctuations
give rise to quantum corrections to the dynamics and this leads to the necessity of a renormalization procedure.
It was shown that while standard General Relativity, possibly together with matter, is not a consistent quantum field theory at perturbative level, i.e. is not perturbatively renormalizable, quadratic gravity is, and can play the role of UV completion of General Relativity. In this case the Einstein-Hilbert term, linear in the curvature, is just a relevant operator which becomes dominant in the low energy, large distance and low curvature regime, whereas the curvature squared terms dominate at Planck energies. Therefore these are expected to play an important role in the very early universe, and possibly also during inflation.

Quadratic gravity is the theory of gravity with action (in signature  $(-,+,+,+)$, natural units
and omitting total derivatives)
\begin{equation}
S=\int d^4x\sqrt{|g|}
\left[\frac{\Mpl^2}{2}(R-2\Lambda)
-\frac{1}{2\lambda}C^2
-\frac{1}{\xi}R^2
\right]\ ,
\end{equation}
where $\Mpl=1/\sqrt{8\pi G_N}$ is the (reduced) Planck mass,
$\Lambda$ is the cosmological constant,
$C_{\mu\nu\rho\sigma}$ is the Weyl tensor.
Quadratic gravity is renormalizable \cite{Stelle1}
but, at least in a naive perturbative analysis in flat space,
leads to pathological propagation.
Indeed, in the spin two sector the propagator is

\begin{equation}
\frac{-4i \lambda }{q^4+\frac12\lambda \Mpl^2 q^2}
=\frac{8 i}{\Mpl^2}\left[\frac{1}{-q^2}-\frac{1}{-q^2-\frac12\lambda \Mpl^2 }\right]\ , \nonumber
\end{equation}
showing the massless graviton and a massive state that is a ghost
and, for $\lambda<0$, also tachyonic.
In the spin zero sector the propagator is
\begin{equation}
\frac{-i \xi/3}{q^4-\frac{1}{12}\xi \Mpl^2 q^2}
=\frac{4 i}{\Mpl^2}\left[-\frac{1}{-q^2}+\frac{1}{-q^2+\frac{1}{12}\xi \Mpl^2 }\right]\ . \nonumber
\end{equation}
The massless pole (a ghost) can be eliminated by gauge transformations,
as in Einstein's General Relativity, whereas the massive state is a tachyon for $\xi>0$.
Thus, to avoid tachyons, we must choose $\lambda>0$ and $\xi<0$.

The theory without the square of the Weyl tensor, which is relevant to the
Starobinsky model, does not have ghosts but is not renormalizable.
Here we shall consider the Starobinsky model as a limit of the renormalizable
quadratic gravity theory, in which the square of the Ricci scalar is in some sense
dominant with respect to the square of the Weyl tensor.
This point of view is motivated by the better quantum properties of the general quadratic gravity theory.
There remains then the problem of the ghosts and tachyons.
There has been some progress recently on both fronts.
Ghosts are generally assumed to imply violation of unitarity,
but several authors have suggested that massive ghosts 
cannot appear as asymptotic states and that the theory is unitary
\cite{Anselmi:2018ibi,Salvio:2018crh,Donoghue:2021cza,Buoninfante:2023ryt}
Instead, there would be a violation of causality at microscopic scales
(scales comparable to the mass of the ghost) and this may be acceptable.
We will not discuss ghosts further and focus instead on the
problem of tachyons, whose presence depends on the signs of
the quadratic curvature terms.
In addition to being renormalizable, quadratic gravity had also been
shown to be asymptotcally free,
but only for a choice of signs that implies the existence of
a massive spin $0$ tachyon, signalling an intrinsically unstable vacuum
\cite{ft1,Avramidi:1985ki}.
A recent recalculation of the beta functions~\cite{Buccio:2024hys}, 
based on scattering amplitudes and including in the running 
the contribution of the infrared logs
(i.e. terms related to a mass threshold like $\log{(E^2/m^2)}$),
leads to a slighly different set of beta functions 
which admit asymptotically free theories lying outside the tachyonic region.
The main aim of this paper is to discuss possible
implications of this result for Starobinsky inflation.

We will briefly review Starobinsky inflation in Section 2
the new beta functions in Section 3 and their implications in Section 4.
Finally, in Section 5 we summarize and conclude with some open questions.

\vskip 1cm
{\it One of us (G.P. V.) had the privilege to collaborate with Alexei: 
``We met several times when he visited our department starting from the early "2000". Our work together was in the context of his stochastic inflation techniques~\cite{stochastic,Starobinsky:1994bd} to define renormalized correlators, extending the results  to the description of renormalized correlators of gauge invariant scalar fluctuations in a quasi de Sitter inflationary phase~\cite{  Finelli:2008zg,Finelli:2010sh}.
I remember him with great pleasure as a very nice person beyond being a great scientist who left us a profound legacy."
}

\section{Starobinsky model and inflation}
\label{sec:2}

Here we shall briefly review the inflationary predictions of the "$R+R^2$" Starobinsky model.
This model has an action containing terms up to second order in the Ricci scalar 
that is usually written in the form
\begin{equation}
S=\frac{\Mpl^2}{2} \int d^4 x \sqrt{-g} \left[ R +\frac{1}{6m^2} R^2 \right] \,.
\end{equation}
Here $m$ is the mass of the propagating spin-zero degree of freedom present in the $R^2$ term. One can appreciate in a simple way the dynamical features of this model in relation to inflation, changing description after moving from the Jordan frame to the Einstein frame. First one introduces a Lagrange multiplier field $X$ to rewrite
\begin{equation}
S=\frac{\Mpl^2}{2} \int d^4 x \sqrt{-g} \left[ \left(1+\frac{X}{3m^2}\right) R -\frac{1}{6m^2} X^2 \right] \,,
\end{equation}
then performs a Weyl transformation $g_{\mu\nu} \to \Omega^2 g_{\mu\nu}$, 
$R \to \Omega^{-2} R - 6 \Omega^-3 \Box \Omega$, 
and in order to obtain a standard Einstein term one chooses $\Omega=\left(1+\frac{X}{3m^2}\right)^{-1/2}$.
The last step, after an integration by parts and neglecting total derivatives, is to define the field $\phi=\sqrt{\frac{3}{2}} \Mpl \ln{\left(1+\frac{X}{3m^2}\right)}$ which has a canonical kinetic term. The rest, written in terms of $\phi$ defines its potential. One obtains
\begin{equation}
S= \int d^4 x \sqrt{-g} \left[ \frac{\Mpl^2}{2} R -\frac{1}{2} \partial_\mu \phi \partial^\mu \phi -V(\phi) \right] \,,
\label{SMEF}
\end{equation}
where $V(\phi)=\frac{3 m^2 \Mpl^2}{4} \left(1-e^{-\sqrt{\frac{2}{3}} \frac{\phi}{\Mpl}}\right)^2$. This is the Starobinsky model in the Einstein frame.

At this point a standard analysis for the study of the FLRW spacetime evolution and gauge invariant fluctuations can be performed~\cite{Mukhanov:1990me, Ellis:2015pla}.
For such a flat potential the slow roll approximation is well suited and as usual one can study the evolution of the scale factor $a(t)$ with Hubble parameter $H=\dot{a}/a$ using the equations: $H^2\simeq V(\phi)/(3\Mpl^2)$ and $3 H \dot{\phi}+V'(\phi) \simeq0$. Using the first equation one can also write the relation for $H$ in terms of the field $\phi$ as
\begin{equation}
H(\phi)=\frac{m}{2} \left(1-e^{-\sqrt{\frac{2}{3}} \frac{\phi}{\Mpl}}\right)\, .
\end{equation}
Clearly the inflationary expansion is of a quasi de Sitter kind, involving in the Einstein frame a potential (which is almost flat) energy density of about $3/4 \,m^2 \Mpl^2$.
Fixing the pivotal momentum scale and then the time of horizon crossing when the corresponding fluctuation freezes, one can define the number of e-folds till the end of inflation $N_e$, and the slow roll parameters $\epsilon_V$ and $\eta_V$ (whose values, when close to one, set the end of the inflationary phase) as
\begin{equation}
N_e=\frac{1}{\Mpl^2}\int^{\phi_*}_{\phi_e} \!\!\!d\phi \frac{V}{V'}\, , \quad \epsilon_V=\frac{\Mpl^2}{2}\left(\frac{V'}{V}\right)^2 , \quad \eta_V=\Mpl^2 \frac{V''}{V} \,.
\end{equation}
One computes the power spectrum of the primordial gauge invariant scalar $P_\zeta$  and tensor (spin $2$) $P_t$  density perturbations which in terms of the spectral tilts around a pivot scale $k_0$ can be written as
\begin{equation}
P_\zeta=A_s \left(\frac{k}{k_0}\right)^{n_s-1}, \quad P_t=A_t \left(\frac{k}{k_0}\right)^{n_t}\,.
\end{equation}
In the lowest approximation the spectral indices can be assumed to be momentum independent (no running). In particular the tilt $n_s$  of the nearly scale invariant scalar power spectrum and the ratio $r=A_t/A_s$ are the most used observables to discriminate between inflationary models. 
They can be computed in the slow roll regime using
\begin{equation}
n_s=1+2\eta_V-6\epsilon_V, \quad r=16 \epsilon_V\ .
\end{equation}
Since tensor perturbations have not been observed yet, there is a strong experimental upper bound $r< 0.036$ at 95\% C.L.~\cite{BICEP:2021xfz}.
From the form of the potential of the Starobinsky model in the Einstein frame in Eq.~\eqref{SMEF} one obtains very simple relations for such observables in terms of the number of e-folds $N_e$:
\begin{equation}
n_s\approx 1-\frac{2}{N_e} , \quad r\approx\frac{12}{N_e^2}  \quad \Rightarrow \quad r\approx3(n_s-1)^2 \,.
\end{equation}
Moreover at horizon crossing the amplitude of the scalar power spectrum can be evaluated in the slow roll approximation to give $A_{s*}= \frac{V^3_*}{12\pi^2 \Mpl^6 {V'}_*^2}$.
The predictions of the quadratic Starobinsky model are in very good agreement with the measurements of CMB data from Planck for $n_s$ (with also $r$ well below the BICEP/KecK experimental constraint) having a central value for the number of e-folds $N_e=56$ and $m/\Mpl={\cal O}(10^{-5})$ to fit the observed values of scalar perturbations.
\section{On the running couplings in quadratic gravity }
\label{sec:3}


Even though gravity is probably the most interesting context 
for higher derivative theories, our interest in the subject started from scalar models.
In particular, a family of scalar theories with $(\phi (-\Box)^k \phi)$ kinetic term
and a shift-invariant $(\partial_\mu\phi\partial^\mu\phi)^2$ interaction
can define conformal field theories in dimensions $d=4(k-1)$ \cite{Safari:2021ocb}.
A special case is the case of $k=2$ and dimension $d=4$, for which the field is dimensionless.
When a standard kinetic term $\partial_\mu\phi\partial^\mu\phi$
is also present, it comes with a mass parameter and the theory
propagates a massless particle and a massive ghost.
It has two free fixed points and one can flow from the
$(\Box\phi)^2$ free theory to the $\partial_\mu\phi\partial^\mu\phi$
free theory, with the dimension of the field changing continuously
from zero to one along the renormalization group trajectory~\cite{Buccio:2022egr}.

This was thought to be a possible toy model for gravity,
with the Hilbert term containing the two-derivative kinetic term
and the quadratic curvature terms containing the four-derivative kinetic term and interactions.
This turned out to be only a poor analogy, for reasons that
we shall mention later, but it spurred a re-examination
of the beta functions of quadratic gravity.

Let us recall here that the main application of the renormalization group in particle physics
is to capture and improve certain properties of scattering amplitudes or correlators.
In (perturbatively) renormalizable theories, the scattering amplitude receives quantum corrections
that in the high energy limit are of the same form as the classical one
and therefore can be subsumed in a change of the coupling
constant with energy.
This energy dependence of the amplitude is logarithmic and
for sufficiently high energies would invalidate perturbation theory.
By considering the infinitesimal change of the coupling with the energy $E$
and integrating it, one effectively resums
certain pieces of the perturbative expansion obtaining an expression
for the amplitude that is valid over a much larger range of energies.
Following \cite{Buccio:2023lzo}, we call the beta functions obtained in this way
the ``physical'' beta functions, as opposed to the ones obtained by taking
derivatives with respect to the parameter $\mu$ of dimensional regularization,
that we shall call the ``mu'' beta functions.

We learn from textbooks that in the limit $E\to\infty$,
these beta functions are the same.
However, this is not universally true for all quantum field theories.
It holds in standard renormalizable quantum field theories,
because in the high energy limit the masses can be neglected
and in the absence of any other dimensionful parameters,
the arguments of the logarithms must necessarily involve the ratio $E/\mu$.
However, in theories with four derivatives, something different can happen.
The propagator behaves typically like $1/(q^4+m^2 q^2)$, 
which improves the convergence
of loop integrals in the UV, but it makes it worse in the IR when we take $m\to 0$.
In particular, depending on details of the interaction,
a one loop integral could behave like $\int \frac{d^4q}{q^4+m^2 q^2}$,
that is logarithmically divergent for $m\to 0$, 
and would also show up in dimensional regularization.
In such cases the mass cannot be neglected and could appear
in the arguments of the logs.
On dimensional grounds, an amplitude could typically have the form
\begin{equation}
{\cal M}(E)=\lambda(\mu)+a\lambda^2(\mu)\log\left(\frac{m^2}{\mu^2}\right)
+b\lambda^2(\mu)\log\left(\frac{E^2}{\mu^2}\right)
+c\lambda^2(\mu)\log\left(\frac{E^2}{m^2}\right)\ ,
\end{equation}
where $\lambda$ is the coupling.
If there are no infrared divergences in the massless limit,
as is the case in four-dimensional theories with standard
two-derivative kinetic terms, $c=a$.
Then, the mass drops out, the amplitude is
\begin{equation}
{\cal M}(E)=\lambda(\mu)+(a+b)\lambda^2(\mu)\log\left(\frac{E^2}{\mu^2}\right)\ ,
\end{equation}
and the ``physical'' beta function is identical to the ``mu'' beta function.
In four-derivative theories with infrared divergences,
$c\not=a$ and taking the derivative with respect to $E$
gives a different result from taking the derivative with respect to $\mu$.
This is the case in quadratic gravity, where infrared divergences appear
if we neglect the Hilbert term.
The Planck mass acts as a IR regulator and its effect is felt
even in the $E\to\infty$ limit.

The explicit calculation leads to the following result for the ``mu" beta functions
\begin{eqnarray}
\beta_\lambda&=&-\frac{1}{(4\pi)^2}\frac{133}{10}\lambda^2\ ,
\label{abl}
\\
\beta_\xi&=&-\frac{1}{(4\pi)^2}\frac{5(72\lambda^2-36\lambda\xi+\xi^2)}{36}\ ,
\label{abx}
\end{eqnarray}
while the ``physical" beta functions are
\begin{eqnarray}
\beta_\lambda&=&-\frac{1}{(4\pi)^2}\frac{(1617\lambda-20\xi)\lambda}{90}\ ,
\label{bdmpl}
\\
\beta_\xi&=&-\frac{1}{(4\pi)^2}\frac{\xi^2-36\lambda\xi-2520\lambda^2}{36}\ .
\label{bdmpx}
\end{eqnarray}
\begin{figure}[t]
\begin{center}
\includegraphics[scale=0.41]{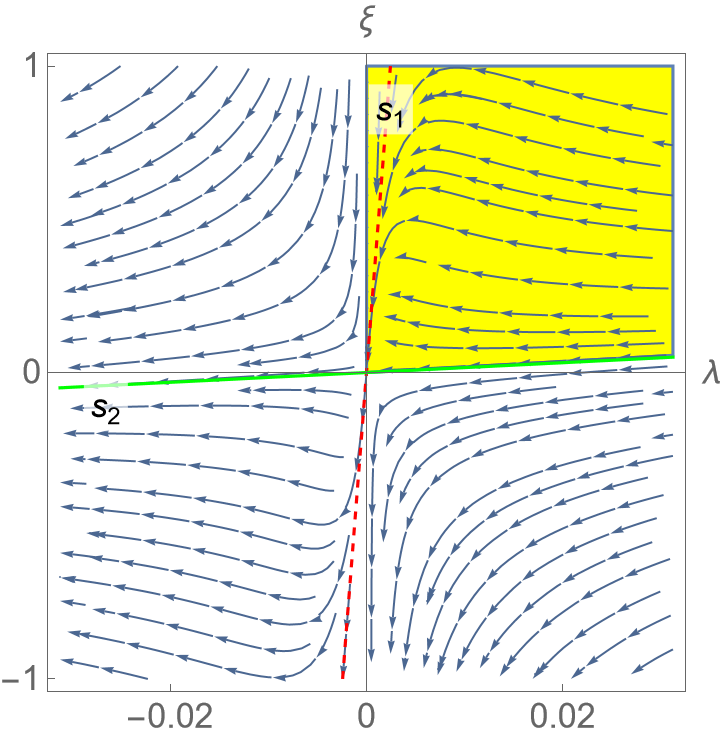}\hspace{1cm}
\includegraphics[scale=0.41]{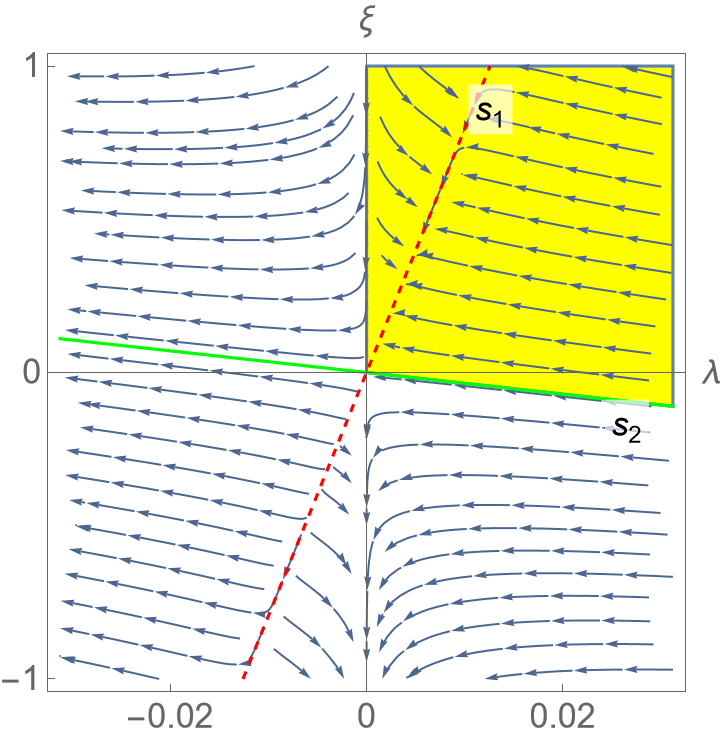}
\caption{Left: old ``mu" flow. Right: new ``physical" flow.}
\label{fig:1} 
\end{center}
\end{figure}
The renormalization group flows defined by these beta functions
are shown in Fig.~\ref{fig:1}.
Their main qualitative features are determined by the presence of a
Gaussian (free) fixed point in the origin, and by
various separatrix curves that delimit different basins of attraction.
In this case the separatrices are loci where the flow is purely radial
and correspond to fixed points for the ratio $\omega=-3\lambda/\xi$.
In the case of the ``mu" flow they are
$$
s_1:\quad \xi=\frac{1291+\sqrt{1637881}}{20}\lambda\approx 128.5\lambda
\quad\Rightarrow \omega=-0.0233
$$
$$
s_2:\quad \xi=\frac{1291 -\sqrt{1637881}}{20}\lambda\approx 0.5601\lambda
\quad\Rightarrow \omega=-5.3558
$$
whereas for the ``physical" flow they are
$$
s_1:\quad \xi=\frac{569+\sqrt {386761}}{15}\lambda\approx 79.4\lambda
\quad\Rightarrow \omega=-0.03778
$$
$$
s_2:\quad \xi=\frac{569-\sqrt {386761}}{15}\lambda\approx -3.53\lambda
\quad\Rightarrow \omega=0.8506
$$

The main new feature to observe is that the basin of attraction of
the Gaussian fixed point in the origin, shown in yellow in both panels of Fig.~\ref{fig:1},
lies entirely in the first quadrant, where the spin zero state is a tachyon,
for the old beta functions but for the new beta functions also extends
to a wedge in the fourth quadrant, where the signs are right to avoid tachyons. 
The spin $2$ massive state in both cases is a ghost (and not a tachyon) for $\lambda>0$,
namely in the first and fourth quadrants.

\section{Discussion}
\label{sec:4}

The Starobinsky inflation is doing very well as an inflationary model.
Pushing the idea that this model should come from a fundamental UV description at high energies, one can ask what happens in the context of a quantum field theory description of gravity, which is perturbatively renormalizable when the lagrangian is quadratic in the curvature. 
Focusing on the one loop renormalized framework of quadratic quantum gravity, which was recalled in the last section, we see that the Starobinsky model, which has no tachyon states is obtained in the limit $\lambda \to \infty$ and $\xi=-\frac{12 m^2}{\Mpl^2}$.

Therefore only in an asymptotically free quadratic gravity theory described by couplings with the ``physical" running encoded in Eqs.~\eqref{bdmpl} and~\eqref{bdmpx} one can find RG renormalized trajectories connecting a theory without tachionic instabilities in the ``low" energy region to the UV Gaussian fixed point.

Pragmatically, one is tempted to consider the region in the theory space of a renormalizable quadratic gravity, 
where the Weyl square $C^2$ term is strongly suppressed with respect to the $R^2$ one, i.e. $\lambda \gg -\xi$ (or $\omega\gg1$) at some suitable energy scale $M$. We could imagine for example $M$ related to the scale of inflation, possibly defined by the potential energy density and therefore of the order $M\approx \sqrt{m \Mpl}$. One might expect these theories to be good candidates to describe a dynamics, in particular an inflationary dynamics, which is very close to the one of the Starobinsky model. Then we observe that a further condition on the couplings must be satisfied: one would require the theory to be free from tachyons at least many order of magnitudes above the scale $M$ which means that we require the RG time $T$ needed for $\xi$ to reach zero in the flow towards UV to be large enough.

Let us discuss how the RG trajectories associated to the ``physical" beta functions are realized.
The special trajectory given by the $s_2$ separatrix is characterized by the relation $\lambda \simeq -0.28 \,\xi$, which is free from tachyons at any scale. Along this special trajectory one has a fixed ratio of the renormalized couplings $\lambda/\xi$, with even a slightly dominant Weyl square term with respect to the $R^2$ one, which is therefore far from the Starobinsky model. Any other trajectory connecting a point with the same negative $\xi$ above this separatrix has a larger value of $\lambda$. In this case the trajectory starting at $(\lambda_i,\xi_i)$, directed towards the UV Gaussian fixed point, crosses the $\lambda$-axis after some positive renormalization group time $T$ and enters the scalar tachyonic region of positive $\xi$, approaches the separatrix $s_1$ and then stays close to it until it reaches the origin after an infinite amount of RG time.
In particular in the case $-\xi_i \ll \lambda_i \ll 1$, one can find a simple relation for the RG time $T$ needed for a negative $\xi$ to flow to the zero value,  since $\lambda$ changes much more slowly than $\xi$, 
which then changes almost linearly with RG time. One finds
\begin{equation}
T\simeq -\frac{\xi_i}{\lambda_i} \frac{1}{\frac{35}{8\pi^2}\lambda_i+\frac{539}{480\pi^2}\xi_i} \simeq -\frac{8\pi^2}{35} \frac{\xi_i}{\lambda_i^2}\,.
\end{equation}
It is also interesting to look at the RG flow trajectory towards the IR starting from this region of parameters.
One can immediately see that an IR Landau pole is typically reached after some finite (negative) RG time $T_{LP}$. 
A good estimate of its position can be be obtained on neglecting the mixed $\xi \lambda$ term in the beta function $\beta_\lambda$ so that one obtains
\begin{equation}
T_{LP}\simeq - \frac{480\pi^2}{539} \frac{1}{\lambda_i}\,,
\end{equation}
which for $\lambda \ll 1$ is very large.

Let us consider a numerical example, considering for an inflationary regime a model at the energy scale $M$ with $\xi \simeq -10^{-9}$ and $\lambda=10^{-5}$, so that the above mentioned suppression factor for the term $C^2$ w.r.t. $R^2$ is $10^{-4}$. In this case one has the RG time $T\simeq 22.6$ for $\xi$ to become zero towards the UV, which means that, following the RG flow, the system would enter the tachyonic region at a scale $M_T=M e^T\simeq 10^{10} M$, while on half a way along the flow since $\xi$ changes almost linearly, one would have $\xi=-0.5 \times 10^{-9}$ at the energy scale $M_{T/2} \approx 10^5 M$.
This shows that the effective theory which approximates the Starobinsky model for the given $C^2$ suppression is valid in a wide interval of energy scales. One can also easily check that the RG flow in the opposite direction towards the IR with the one loop ``physical'' beta functions indeed flows to hit a Landau pole, but located at extreme low energy scale. In the numerical example given above  this would happen at an RG time $T_{LP}\simeq - 8.6\times 10^5$, which is beyond any meaningful physical scale, corresponding to lengths much larger that the extension of the known universe. Instead for a much shorter flow towards the IR, for example of $T=-20$ (corresponding to an energy scale $e^T$ smaller, i.e. about  $9$ order of magnitude smaller) one finds a  small change in the value of $\xi$, with $\xi=-1.88 \times 10^{-9}$, and $\lambda$ practically unchanged, which means that there is a wide range of energy scales with slowly changing parameters of the model.

\section{Conclusions}
\label{sec:5}
We have discussed how the Starobinsky model, which makes very succesful inflationary predictions, could be thought of as an approximation to a renormalizable and UV complete quadratic gravity theory. Then, the use of the so called ``physical'' scheme for the running couplings
leads to an extension, free of tachionic instabilities, of the domain of validity of the model over a wide range of energy scales along the RG trajectory, both towards the UV and the IR.

Nevertheless within this picture there are several aspects that deserve to be investigated
in greater detail. 
The homogeneous scale factor evolution of the FLRW space-time is not affected by the presence of the conformally invariant Weyl square term but certainly at the level of fluctuations we expect non trivial effects, possibly both of classical and quantum origin. 
Therefore one has to check that the differences from the Starobinsky model are not spoiling its predictions. 
In a very recent work for example the spectral tilt and amplitude of tensorial power spectrum has been computed in the presence of the spin $2$ ghost states and shown to be slightly suppressed~\cite{Kubo:2025jla}. 
On the other hand recently a linear classical perturbative analysis of scalar metric gauge invariant perturbations in presence of the Weyl square term has shown an exponential grow and oscillating behavour~\cite{DeFelice:2023psw}. This might signal an instability which should be further investigated, eventually performing a non linear classical analysis. Indeed there are known examples of dynamical higher derivative systems which one would naively expect to be unstable and which, instead, show within regions of initial condition to have a bounded evolution~\cite{Deffayet:2023wdg}, a behavior completely unseen by a linear analysis. In a recent numerical analysis of quadratic gravity evolution~\cite{Held:2023aap}, which focus on metric perturbations around a black hole, it has been shown that while at linear level instabilities can be encountered, the full non linear evolution is physically stable. Therefore further investigations are needed in the context of inflationary dynamics. Moreover one should later address also the quantum aspect with the treatment of the ghost states.

There are also some more theoretical aspects that require further analysis.
One might wonder if the ``physical'' beta functions considered here are the relevant way to define a perturbative renormalized model to be used in inflationary studies. These have been derived in the context of on shell scattering amplitudes (on an almost flat background), that is are based on an analysis of in-out correlators of the theory. Normally in cosmology one deals with in-in correlators. There have been recently some analysis suggested by the so called cosmological bootstrap approach, where it has been shown that one can deduce in-in correlators from in-out ones~\cite{Donath:2024utn}. But this equivalence has been derived in a context where several assumptions have been made: a maximally symmetric spacetime background  of de Sitter type is considered, unitarity of the theory at hands was satisfied, and IR divergences were absent.
It is therefore important to understand up to which point this correspondence can be further pushed.

\vskip 1cm
\noindent
{\bf Acknowledgements}

G.P.V. thanks Michele Cicoli and R.P. thanks Luca Buoninfante for useful discussions.

\vskip 1cm


\end{document}